\documentclass[12pt,letterpaper]{article}
\usepackage[latin1]{inputenc}
\usepackage{amsmath}
\usepackage{mathptmx,courier,textcomp}
\usepackage{amsfonts}
\usepackage{amssymb}
\usepackage{cite}
\usepackage[small]{caption} 
\usepackage{graphicx}
\usepackage{epstopdf}
\usepackage{ulem}
\usepackage[top=1in,bottom=1in,left=1in,right=1in]{geometry}
\title{Time- and frequency-resolved dynamics of resonant two-photon terahertz quantum cascade lasers}

\author{Muhammad Anisuzzaman Talukder\\
\small{Department of Electrical and Electronic Engineering}\\
\small{Bangladesh University of Engineering and Technology}\\
\small{Dhaka 1205, Bangladesh}\\
\small{\it anis@eee.buet.ac.bd}}
\date{}
\begin{document}
\maketitle
%-------------------------------- 
\begin{abstract}
Two-photon terahertz (THz) quantum cascade lasers (QCLs) can immensely improve the conventional applications of THz frequency sources and open windows to new applications due to their ability to generate quantum-entangled twin photon beams. The rich intrinsic non-linearity in a two-photon THz QCL arises from cascaded photon transitions, leading to time-resolved and frequency-resolved dynamics that are less predictable. This work investigates the time- and frequency-resolved dynamics of a resonant two-photon THz QCL by numerically solving the coupled Maxwell-Bloch equations. Our findings indicate that the output intensity is significantly higher, and the emission spectra are much broader in a two-photon THz QCL than in a conventional one-photon THz QCL. Additionally, a pronounced Risken-Nummedal-Graham-Haken (RNGH) instability onsets for a resonant two-photon QCL at a smaller pumping current. We also observe an increase in the number of lasing modes, as well as a higher power per line in a resonant two-photon QCL. Furthermore, Rabi splitting is observed in the emission spectra at high pumping currents.
\end{abstract}
%--------------------------------
%
%%%%%%%%%%%%%%%%%%%%%%%%%%  body  %%%%%%%%%%%%%%%%%%%%%%%%%%
%-----------------------------------------------------------------------------------------------------------------------------------------------------------------
%
\section{Introduction}
The realization of lasers with two-photon emissions for each excited state relaxation has been pursued since the 1960s \cite{sorokin64,prokhorov65}. Proposals of two-photon lasers were followed by substantial theoretical efforts over the next three decades foreseeing the potential of such lasers \cite{bunkin66,schlemmer80,lewenstein90,zakrzewski91a,zakrzewski91b,zakrzewski91c}. A two-photon laser is unique in its properties as a non-classical light source \cite{boone89,boone90} and offers many exciting applications, such as precision measurement, quantum gyroscope, quantum communication, and generation of polarization-entangled twin beams, in addition to many conventional frequency-specific applications of lasers \cite{yuen78,pfister01}. Unfortunately, the potential of two-photon lasers remains unexplored as their realization is still challenging. A natural material that can be used as the gain medium with an appropriate electronic structure for two cascaded photon emissions is hard to find. Therefore, the successes in realizing two-photon lasers have been limited---the first demonstration was in 1987 using rubidium atoms and working in the microwave frequency range \cite{brune87b}, and the second and the last demonstration was in 1992 using barium atoms and working in the optical frequency range \cite{gauthier92}.

Lasers based on semiconductor materials are advantageous for compactness, efficiency, and ease of use. Prospects of two-photon lasers based on interband transitions \cite{ironside92,marti03} and intersubband transitions \cite{ning04} in semiconductor materials have been theoretically investigated. Recently, there has been a report of interband two-photon spontaneous emission from semiconductors \cite{hayat08}. However, a semiconductor two-photon laser is yet to be realized, mainly due to a lack of suitable structure. While semiconductor technology based on interband transition is more mature than that based on intersubband transition, semiconductor technology based on intersubband transition offers more design flexibility than that based on interband transition. Intersubband transitions also offer a dipole moment $\sim$10 times greater than that of interband transitions \cite{ning04}. 

The intersubband transitions in the conduction band of quantum heterostructures are wisely used in quantum cascade lasers (QCLs) \cite{faist94}. The intersubband transition energies in a quantum heterostructure depend on the layer thicknesses of two alternating semiconductor materials and, therefore, can be varied merely by changing the layer thicknesses. QCLs have been demonstrated over a broad spectral range and are a mature technology in both mid-infrared (mid-IR) and terahertz (THz) frequency ranges \cite{botez23}. THz QCLs are critical light sources for their applications in biomedical sensing and imaging, molecular spectroscopy, astronomy, and security screening \cite{vitiello22}. Since their first demonstration in 2002 \cite{kohler02}, THz QCLs have come a long way in achieving high power and continuous wave operation \cite{williams07,kumar11}. Band-structure engineering and cavity designs have enabled THz QCLs to achieve modelocked pulses and stable frequency combs \cite{burghoff14,barbieri11,wang17}. However, practical and commercial applications of THz QCLs are still significantly limited by their output power and narrow emission spectra.

Recently, THz QCLs have been proposed as an ideal tool for realizing a semiconductor laser based on two-photon emissions \cite{talukder22}. In a two-photon THz QCL, there would be two cascaded photon transitions in the active region due to the relaxation of each excited electron. Analytical solutions of Maxwell-Bloch equations have shown that resonant two-photon THz QCLs will be more efficient and emit greater output power than conventional one-photon THz QCLs. It has been predicted that they will be broadband and can be made widely tunable. Thus, two-photon THz QCLs will improve on the conventional applications, e.g., they can be used to make frequency combs with greater dynamic range and more power per line. Many more exciting applications are possible, such as high-precision measurements, laser gyroscope, quantum communication, and generating polarization-entangled twin beams \cite{yuen78,pfister01}.

This work investigates two-photon THz QCLs by solving coupled Maxwell-Bloch equations using the finite-difference time-domain (FDTD) technique. We specifically examine the time- and frequency-resolved characteristics of resonant two-photon THz QCLs. As predicted by analytical calculations, the output emission intensity is much higher, and emission spectra are much broader compared to conventional QCLs. We also investigate the chaotic emission profile of a two-photon THz QCL, which arises due to the onset of Risken-Nummedal-Graham-Haken (RNGH) instability. Notably, this instability starts with a lower carrier density in a two-photon THz QCL than in a conventional one-photon THz QCL. Furthermore, we observe Rabi flopping in the carrier density and Rabi splitting at the emission spectra as the carrier density increases.

%---------------------------------------------
%\section*{Theoretical model}
\section{Maxwell-Bloch equations}
In the proposed two-photon THz QCL of Ref.~\citeonline{talukder22}, two resonant gain stages are cascaded in the active region. The upper energy level of the first gain stage acts as the lower energy level of the second gain stage in the cascade. Therefore, the middle level in a three-level system is shared by both the photon transitions. In a three-level system, the first gain stage is between the lower level 1 and the intermediate level 2. The second gain stage is between the intermediate level 2 and the upper level 3. The transition frequency in the two gain stages is resonant.

To develop a mathematical model of the proposed resonant two-photon THz QCL, we denote the first and second gain stages as `a' and `b', respectively. We assume that levels 1 and 2 in the gain stage `a' are coupled by a dipole moment $d_a$ that decays in a characteristic dephasing time $T_{2a}$. Similarly, levels 2 and 3 in gain stage `b' are coupled by a dipole moment $d_b$ that decays in a characteristic dephasing time $T_{2b}$. The polarization densities $\eta_a$ and $\eta_b$ in gain stages `a' and `b' can be given by Bloch equations \cite{valcarcel95,talukder14,talukder09apl}
%
%-------------------------
% EQUATION 1
\begin{subequations}
\begin{align}
% Polarization Equation for gain stage 'a'
\frac{\partial\eta_{a\pm}}{\partial t}&=
\frac{id_{a}}{2\hslash}\left[(\rho_{2}-\rho_{1})E_\pm + (\rho_{2}^{\mp}-\rho_{1}^{\mp})E_\mp\right] - \frac{\eta_{a\pm}}{T_{2a}},\\
%
% Polarization Equation for pump pulse
\frac{\partial\eta_{b\pm}}{\partial t}&=
\frac{id_b}{2\hslash}\left[(\rho_{3}-\rho_{2})E_\pm + (\rho_{3}^{\mp}-\rho_{2}^{\mp})E_\mp\right] - \frac{\eta_{b\pm}}{T_{2b}},%\label{eq:mbe_pump_pol_gain}
\end{align}
\end{subequations}
where the variable $E$ is the envelope of the electric field, $\rho_x$ is the population density in level $x$, $\rho_x^\pm$ is the grating created on the population density in level $x$ due to the interaction between the forward and backward traveling electric fields, and $\hslash$ is the Planck constant. We use a subscript $+$ ($-$) to denote a quantity related to the field propagating in the positive (negative) $z$-direction. In addition, Eq.~(1) assumes that both transitions are resonant and they emit photons of the same energy, which is essential to achieve powerful lasing with rich non-linearity.

The change in carrier densities due to emission stimulated by a propagating electric field and non-radiative relaxation can be described by \cite{valcarcel95,talukder09pra,talukder14,talukder09apl}
%-------------------------
\begin{subequations}
\begin{align}
% Population Equation \rho_1
\frac{\partial\rho_{1}}{\partial t}&=
-\lambda - \frac{id_a}{2\hslash}(E^*_+\eta_{a+} + E^*_-\eta_{a-} - {\rm c.c.}) + \frac{\rho_{2}}{\tau_2},\label{eq:mbe_rho_1}\\
%
% Population Equation \rho_2
\frac{\partial\rho_{2}}{\partial t}&=
- \frac{id_b}{2\hslash}(E^*_+\eta_{b+} + E^*_-\eta_{b-} - {\rm c.c.}) + \frac{id_{a}}{2\hslash}(E^*_+\eta_{a+} + E^*_-\eta_{a-} - {\rm c.c.}) - \frac{\rho_{2}}{\tau_2} + \frac{\rho_{3}}{\tau_3},\label{eq:mbe_rho_2}\\
%
% Population Equation \rho_3
\frac{\partial\rho_{3}}{\partial t}&=
\lambda + \frac{id_b}{2\hslash}(E^*_+\eta_{b+} + E^*_-\eta_{b-} - {\rm c.c.}) - \frac{\rho_{3}}{\tau_3},\label{eq:mbe_rho_3}\\
%
% Population Inversion Grating Equation \rho_{12}^{\pm}
\frac{\partial\rho_{1}^\pm}{\partial t}&=
-\frac{id_a}{2\hslash}(E^*_\pm\eta_{a\mp} - E_\mp\eta^*_{a\pm} ) - \frac{\rho_{1}^\pm}{\tau_1},\label{eq:mbe_rho_12}\\
%
% Population Inversion Grating Equation \rho_{2}^{\pm}
\frac{\partial\rho_{2}^\pm}{\partial t}&=
-\frac{id_b}{2\hslash}(E^*_\pm\eta_{b\mp}  - E_\mp\eta^*_{b\pm} ) + \frac{id_a}{2\hslash}(E^*_\pm\eta_{a\mp}  - E_\mp\eta^*_{a\pm} ) - \frac{\rho_{2}^\pm}{\tau_2} +\frac{\rho_{3}^\pm}{\tau_3}, \label{eq:mbe_rho_22}\\
%
% Population Inversion Grating Equation \rho_{3}^{\pm}
\frac{\partial\rho_{3}^\pm}{\partial t}&=
\frac{id_b}{2\hslash}(E^*_\pm\eta_{b\mp} - E_\mp\eta^*_{b\pm} ) - \frac{\rho_{3}^\pm}{\tau_3},\label{eq:mbe_rho_32}
\end{align}
\end{subequations}
%---------------------------
where the parameter $\lambda$ is the injection or pumping rate to level 3, and the extraction rate from level 1 and $\tau_x$ is the lifetime of level $x$. The two-photon QCL structure operates under a fixed bias, as discussed in Ref.~28. The population densities in Eq.~(2) remain at their equilibrium values, provided that the electric fields do not interact with the two-photon gain medium. Since the applied bias remains constant, the transition frequencies of the photons are also assumed to be fixed as the population densities change according to Eq.~(2).

The evolution of the electric field can be given by Maxwell's equation \cite{valcarcel95,talukder14,talukder09apl}
\begin{equation}
% Maxwell Equation
\frac{n}{c}\frac{\partial E_\pm}{\partial t}=
\mp \frac{\partial E_\pm}{\partial z} - i\frac{N_a \Gamma d_a k}{\epsilon_0 n^2}\eta_{a\pm} - i\frac{N_b \Gamma d_b k}{\epsilon_0 n^2}\eta_{b\pm} - lE_{\pm},\label{eq:mbe_pump_efield_gain}\\
\end{equation}
where $n$ denotes the index of refraction, $c$ denotes the speed of light, $\epsilon_0$ denotes the vacuum permittivity, $N_a$ and $N_b$ denote the electron densities in gain stages `$a$' and `$b$,' respectively, $\Gamma$ denotes the overlap factor between the laser mode and the active region, $l$ denotes the linear cavity loss per unit length not including mirror losses, and $k$ denotes the wave number associated with the THz radiation. The carrier densities $N_a$ and $N_b$ are related to the total carrier density in a period $N$ by $N_a=N(\rho_2-\rho_1)$ and $N_b=N(\rho_3-\rho_2)$.

%--------------------------------------------------------------------------------------------------------------------------------------
%--------------------------------------------------------------------------------------------------------------------------------------
\section{Simulation results} 
We numerically solve the coupled Eqs.~(1)--(3) using the FDTD technique to calculate the time- and frequency-resolved characteristics of a two-photon THz QCL. The electric field components $E_+$ and $E_-$ travel in opposite directions in the cavity of the three-level gain medium. The initial conditions for $E_+$ and $E_-$ are set by a random noise created by spontaneous emission. The traveling electric field components are reflected by the cavity edges with a coefficient $r_1 = r_2 = (n - n_a)/(n + n_a)$, where $r_1$ and $r_2$ are the reflectivities at the left and right edges, respectively, and $n$ and $n_a$ are the indices of refraction of the gain medium and air. The output electric field ($E_{\rm out}$) is calculated from $E_+$ at the right facet as $E_{\rm out} = \sqrt{1-r^2_2}~E_+$. The frequency-resolved output characteristics are studied by calculating the Fourier transform of $E_{\rm out}$ recorded over 200 round-trip time. While the energy-resolved carrier densities, such as $\rho_2$ and $\rho_3$, are obtained from the solutions of coupled Maxwell-Bloch equations, the population inversions of gain stages `a' ($\Delta_a$) and `b' ($\Delta_b$) are calculated as $\Delta_a=(\rho_2-\rho_1)/(\rho_1+\rho_2)$ and $\Delta_b = (\rho_3-\rho_2)/(\rho_2+\rho_3)$.

This work investigates the time- and frequency-resolved evolutions of lasing in a resonant two-photon THz QCL. We solve the Maxwell-Bloch equations for 200 cavity round-trip times with an initial condition set by the spontaneous emission. We study the start of lasing from a random noise due to spontaneous emission, the rapid rise of light intensity due to two-photon emissions from each electronic relaxation, the onset of self-pulsations due to multimode RNGH instability, the quasi-periodic steady-state intensity fluctuations, and the broadening of the emission spectrum. In FDTD simulations, we use the parameter values given in Table 1. The parameters that depend on the band-structure design are taken from the two-photon THz QCL presented in Ref.~\citeonline{talukder22}. Other cavity and dimensional parameters are chosen from standard parameter regimes of THz QCLs. However, we vary the key parameters, such as the dipole moment $d_b$ and the total carrier density $N$, to understand how the light intensity and the emission spectrum are affected due to two cascaded photon transitions in a two-photon THz QCL. The studied in-depth cavity dynamics will help design and optimize a two-photon THz QCL and understand the characteristics of a conventional one-photon THz QCL.
%
%-----------------------------------
% Table 1
\begin{table}[htbp]
\caption{Key parameter values of the resonant two-photon QCL.}\label{table1}
\centering
\begin{tabular}[p{}]{l  c  c }
\hline\hline Parameter & Symbol & Value\\
\hline 
Lifetime & $\tau_1$ & 0.5 ps\\
Lifetime & $\tau_2$, $\tau_3$ & 50 ps\\
Coherence time & $T_{2a}$, $T_{2b}$ & 1 ps\\
Transition frequency & $f$ & 2 THz\\
Cavity length & $L_c$ & 2 mm\\
Linear loss & $l$ & 1 mm$^{-1}$\\
%Facet reflectivity & $r_1$, $r_2$ & 0.53\\
Refractive index & $n$ & 3.3\\
Diffusion coefficient & $D$ & 4600$\times$10$^{-12}$ mm$^2$/ps\\
\hline\hline
\end{tabular}
\end{table}
%
%--------------------------------------------------------------------------------
%--------------------------------------------------------------------------------
%------------------------------------
% Fig.~1
\begin{figure}[htb]
\centering
\includegraphics[width=3.6in,keepaspectratio]{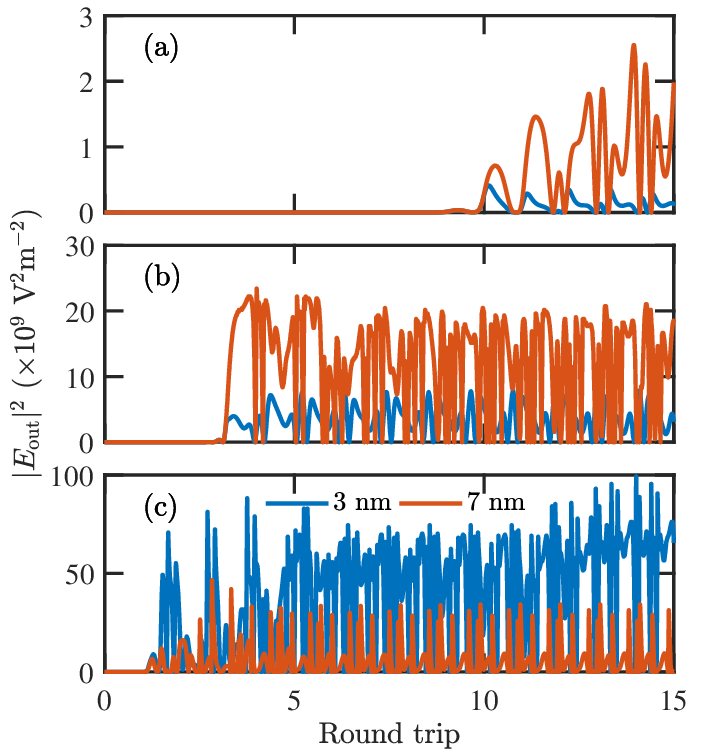}
\caption{Output intensity of a two-photon THz QCL during the first 15 cavity round-trip time for $d_b = 3$ and 7 nm with (a) $N=10^{15}$ cm$^{-3}$, (b) $N=2\times 10^{15}$ cm$^{-3}$, and (c) $N=5\times 10^{15}$ cm$^{-3}$. In each case, $d_a=7$ nm.}\label{Fig_1}
\end{figure}
%-----------------------------------
%---------------------------
\subsection{Start of lasing}  
We investigate a two-photon THz QCL's reaching threshold and the transient dynamics as it starts lasing from a random quantum noise. We vary $N$ to find the threshold carrier density needed for lasing. The lasing starts at $N=0.7\times 10^{15}$ cm$^{-3}$. The threshold condition does not vary with $d_b$ in a two-photon THz QCL. In Fig.~1, we show $|E_{\rm out}|^2$ during the first 15 round-trip time obtained from FDTD simulations for different values of $d_b$ and $N$. We vary the dipole moment $d_b$ to 3 and 7 nm for three different carrier densities: (a) $N=10^{15}$ cm$^{-3}$, (b) $N=2\times 10^{15}$ cm$^{-3}$, and (c) $N=5\times 10^{15}$ cm$^{-3}$. Notably, the ranges in which $d_b$ and $N$ are varied are typically obtained in THz QCLs \cite{williams03, kumar08, liu12}. We observe that the laser starts lasing within the first 15 round-trip time in each case and starts early as $N$ increases. However, for a fixed $N$, the laser starts simultaneously for both $d_b$. The independence of the lasing threshold and lasing start time from $d_b$ represents that the laser starts from a transition from level 2 to level 1. However, once the laser starts lasing due to a transition from level 2 to 1, resonant photons are also created due to transitions from level 3 to level 2. 
%
%----------------------------
% Fig.~2
\begin{figure}[htbp]
\centering
\includegraphics[width=3.6in,keepaspectratio]{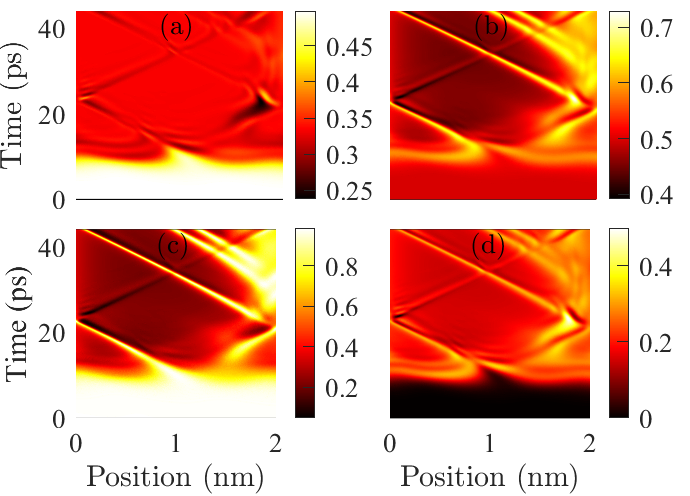}
\caption{Carrier densities and population inversions (a) $\rho_2$, (b) $\rho_3$, (c) $\Delta_a$, and (d) $\Delta_b$ in the two-photon THz QCL during the 2nd cavity round-trip time when $d_a=7$ nm, $d_b=7$ nm and $N=5\times10^{15}$ cm$^{-3}$.}\label{Fig_2}
\end{figure}
%-------------------------------

Figure 1(c) shows that the laser reaches the threshold and starts lasing during the second round-trip time when $d_a=d_b=7$ nm and $N=5\times 10^{15}$ cm$^{-3}$. In Fig.~2, we show the time evolution of $\rho_2$, $\rho_3$, $\Delta_a$, and $\Delta_b$ during the second round-trip time when $d_a=d_b=7$ nm and $N=5\times 10^{15}$ cm$^{-3}$. The change in population densities and inversions when the laser reaches the threshold and starts to lase is dramatic and gives insights into the dynamics of a two-photon THz QCL. We observe that $\rho_2$, $\rho_3$, $\Delta_a$, and $\Delta_b$ remain at their equilibrium values before the lasing starts. Since the lasing starts from the gain transitions from level 2 to level 1, $\rho_2$ depletes to level 1 before a change occurs in $\rho_3$. However, almost instantaneously, $\Delta_b$ starts to grow, and $\rho_3$ starts to deplete to level 2.

%-------------------------------
%Fig.~3
\begin{figure}[t]
\centering
\includegraphics[width=3.6in,keepaspectratio]{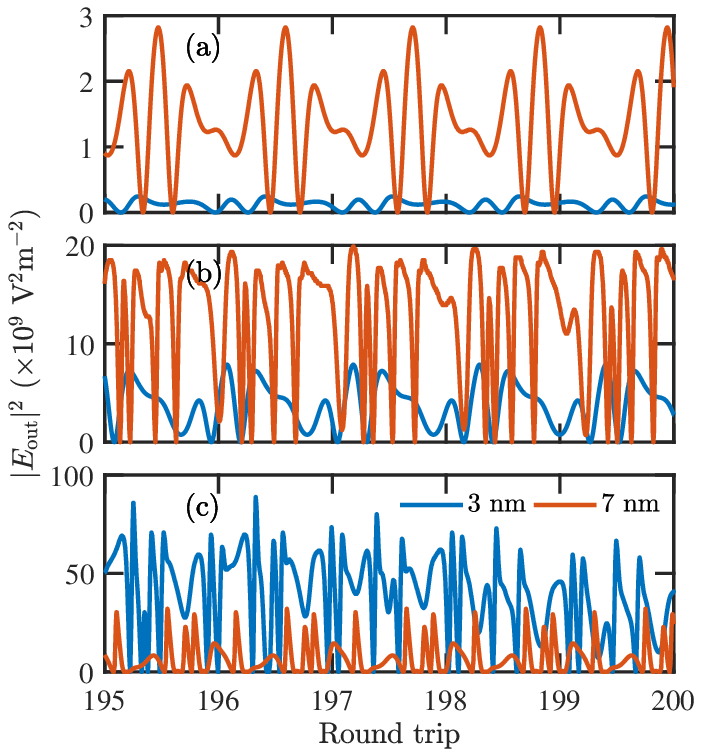}
\caption{Output intensity ($|E_{\rm out}|^2$) of a two-photon THz QCL during last 5 round-trip time of 200 round-trip time simulation for $d_a=3$ and 7 nm with (a) $N=10^{15}$ cm$^{-3}$, (b) $N=2\times 10^{15}$ cm$^{-3}$, and (c) $N=5\times 10^{15}$ cm$^{-3}$. In each case, $d_b=7$ nm.}\label{Fig_3}
\end{figure}
%------------------------------
%
\subsection{Quasi steady-state}
In Fig.~3, we show $|E_{\rm out}|^2$ during the last 5 round-trip time of the 200-round-trip-time FDTD simulations for $d_b=3$ and 7 nm and for the carrier densities (a) $N=10^{15}$ cm$^{-3}$, (b) $N=2\times 10^{15}$ cm$^{-3}$, and (c) $N=5\times 10^{15}$ cm$^{-3}$. Although we show $|E_{\rm out}|^2$ only for five cavity round-trip time, we observe similar behavior once the lasing starts and the initial transient behavior settles within a few round trips. We note that $|E_{\rm out}|^2$ consists of quasi-steady-state periodic oscillations. The two-photon transition is stronger when $d_b=7$ nm than $d_b=3$ nm. The additional photon transition in a two-photon THz QCL due to non-zero $d_b$ helps to increase the photon density in two ways: First, resonant photon transitions from level 3 and 2 add to those from level 2 to 1, and second, the photon-assisted transition from level 3 feeds electrons to level 2 at a fast rate. Therefore, in Figs.~3(a) and (b), $|E_{\rm out}|^2$ is much greater when $d_b=7$ nm than when $d_b=3$ nm. We also note that $|E_{\rm out}|^2$ increases significantly when $N$ increases. However, due to the intense electric field inside a two-photon THz QCL's cavity, the gain saturates for a higher pumping rate. Therefore, the output intensity increases more with $d_b=3$ nm than with $d_b=7$ nm when carrier density increases. 

When the carrier density increases to $5\times 10^{15}$ cm$^{-3}$, the $|E_{\rm out}|^2$ profile significantly changes for $d_b=7$ nm, as shown in Fig.~3(c). Although $|E_{\rm out}|^2$ is quasi-periodic, the light intensity is concentrated in sharp pulses in both time and space. Such a self-pulsating behavior manifests RNGH instability due to broad gain bandwidth over multiple axial modes. The RNGH instability excites the multimode regime in lasers and creates a chaotic intensity profile in the time domain that consists of self-pulsations \cite{risken68}. The threshold for RNGH instability in QCLs is less than most other lasers, and the RNGH instability significantly affects the time- and frequency-resolved characteristics of QCLs \cite{gordon08,talukder09prl}. In a two-photon THz QCL, we observe formations of pulses when the pumping increases. The decrease in intensity when increased pulsations are observed can be explained using the pulse area theorem \cite{talukder09pra}.
%
%---------------------------
% Fig.~4
\begin{figure}[htbp]
\centering
\includegraphics[width=3.6in,keepaspectratio]{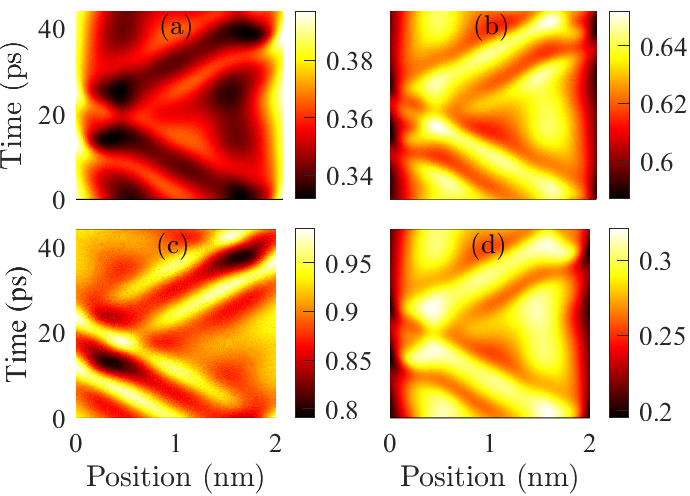}
\caption{Carrier densities (a) $\rho_2$ and (b) $\rho_3$ and population inversions (c) $\Delta_a$ and (d) $\Delta_b$ in the two-photon THz QCL during the 200th cavity round-trip time when $d_a=d_b=7$ nm and $N=10^{15}$ cm$^{-3}$.}\label{Fig_4}
\end{figure}
%
%--------------------------------------------------------------------------
%\subsubsection*{Carrier density}

To understand the change in $|E_{\rm out}|^2$, we investigate the energy-resolved carrier densities $\rho_2$ and $\rho_3$, and the population inversions $\Delta_a$ and $\Delta_b$ when $d_b$ and $N$ change. In Fig.~4, we show the time evolution of $\rho_2$, $\rho_3$, $\Delta_a$, and $\Delta_b$ during the 200th round-trip time when $d_a=d_b=7$ nm, and $N=10^{15}$ cm$^{-3}$. In Fig.~5, we show the time evolution of the same parameters during the 200th round-trip time when $d_a = d_b = 7$ nm and $N=5\times 10^{15}$ cm$^{-3}$. The changes in the 200th round-trip time represent the quasi-steady-state changes inside the cavity. We note that pulses are formed with increased light intensity at a greater $N$ due to RNGH instability. As the pulses propagate, they strongly perturb the carrier densities and the population inversions. The dark-shaded regions represent intense light pulses, while a uniform color means a background light intensity. In contrast, a lighter-shaded region represents a lack of light intensity.

We note that $\rho_2$ does not change much in both Figs.~4 and 5. However, although $\rho_3$ does not change when $N=10^{15}$ cm$^{-3}$, as shown in Fig.~4, $\rho_3$ significantly changes when $N=5\times 10^{15}$ cm$^{-3}$, as shown in Fig.~5. The significant change in $\rho_3$ is due to the propagation of intense pulses in the cavity. When a pulse propagates in the cavity, the population of level 2, $\rho_2$, depletes to level 1, and simultaneously, the population of level 3, $\rho_3$, depletes to level 2. Therefore, $\rho_2$ changes less. We note that the background $\rho_2$ and $\rho_3$ values are $\sim$0.35 and $\sim$0.5 in both Figs.~4 and 5 that are close to the quasi-steady-state values calculated using the analytical formulation \cite{talukder22}. We note that $\Delta_a$ is stronger than $\Delta_b$. Level 1 has a non-radiative scattering lifetime of only $\sim$0.5 ps. Therefore, level 1 depletes fast to level 3 of the next period of the QCL structure. Consequently, $\Delta_a$ is usually much greater than $\Delta_b$ except when light intensity grows too intense in the cavity.
%
%--------------------------------
% Fig.~5
\begin{figure}[htbp]
\centering
\includegraphics[width=3.6in,keepaspectratio]{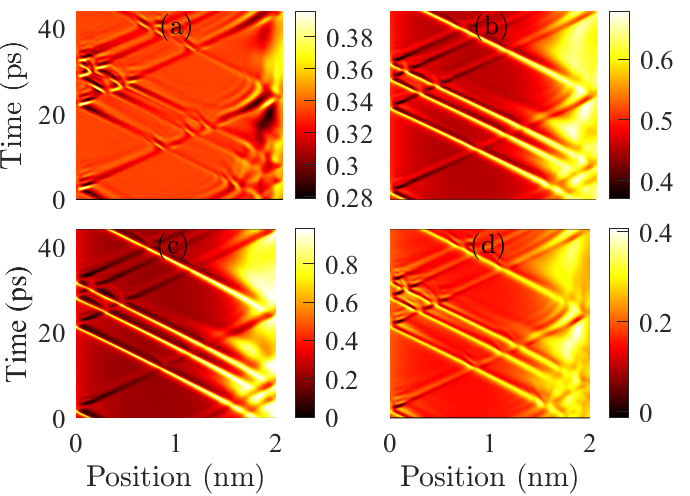}
\caption{Carrier densities (a) $\rho_2$ and (b) $\rho_3$ and population inversions (c) $\Delta_a$ and (d) $\Delta_b$ in the two-photon THz QCL during the 200th cavity round-trip time when $d_a=d_b=7$ nm and $N=5\times10^{15}$ cm$^{-3}$.}\label{Fig_5}
\end{figure}
%-------------------------------
%--------------------------------
% Fig.~6
\begin{figure}[h!]
\centering
\includegraphics[width=3.6in,keepaspectratio]{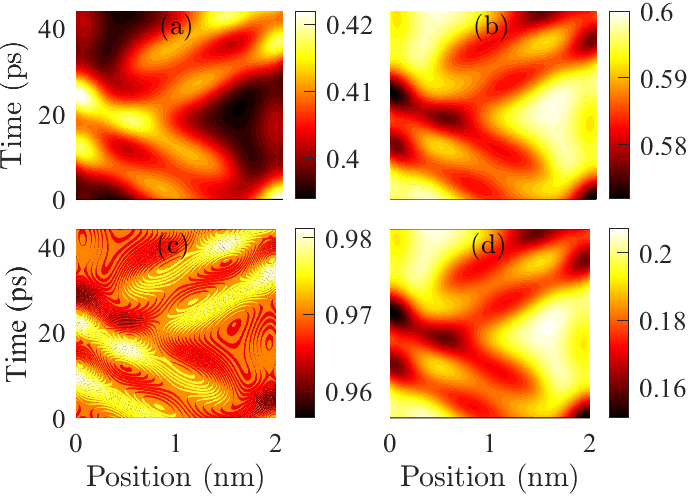}
\caption{Carrier densities (a) $\rho_2$ and (b) $\rho_3$ and population inversions (c) $\Delta_a$ and (d) $\Delta_b$ in the two-photon THz QCL during the 200th cavity round-trip time when $d_a=7$ nm, $d_b=3$, nm and $N=10^{15}$ cm$^{-3}$.}\label{Fig_6}
\end{figure}
%------------------------------------

In Figs.~6 and 7, we show $\rho_2$, $\rho_3$, $\Delta_a$, and $\Delta_b$ when $d_a=7$ nm and $d_b=3$ nm with $N=10^{15}$ cm$^{-3}$ and $N=5\times 10^{15}$ cm$^{-2}$, respectively. We note that $\rho_2$, $\rho_3$, $\Delta_a$, and $\Delta_b$ change less when $d_b$ decreases to 3 nm with $N=10^{15}$ cm$^{-2}$. The two-photon nature of the laser is less exhibited compared to the results presented with $d_b=7$ nm. However, with $N=5\times 10^{15}$ cm$^{-2}$, the changes in carrier densities and population inversions are similar to that with $d_b=7$ nm due to the enhanced photon density within the cavity and photon-induced rapid carrier transitions between the energy levels.

%-----------------------------------
% Fig.~7
\begin{figure}[ht]
\centering
\includegraphics[width=3.6in,keepaspectratio]{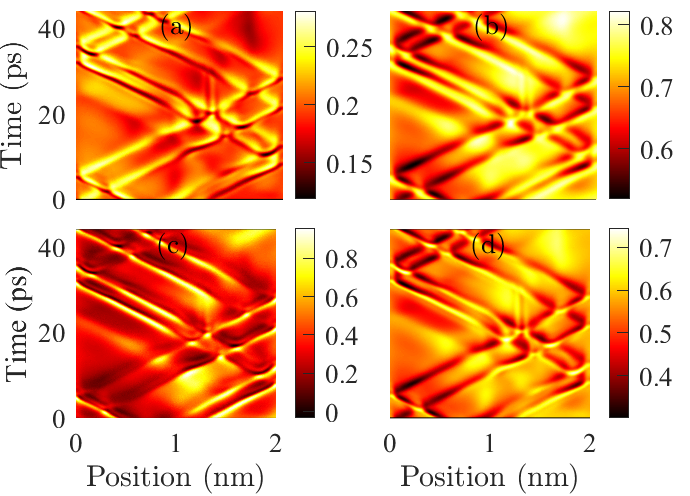}
\caption{Carrier densities (a) $\rho_2$ and (b) $\rho_3$ and population inversions (c) $\Delta_a$ and (d) $\Delta_b$ in the two-photon THz QCL during the 200th cavity round-trip time when $d_a=7$ nm, $d_b=3$ nm, and $N=5\times10^{15}$ cm$^{-3}$.}\label{Fig_7}
\end{figure}
%
%---------------------------------------------------------------------------
%----------------------------------
% Fig.~8
\begin{figure}[ht!]
\centering
\includegraphics[width=3.6in,keepaspectratio]{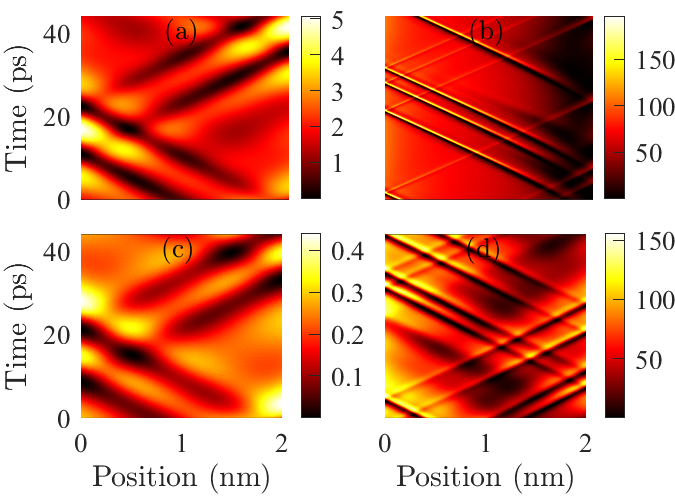}
\caption{Light intensity ($|E_1|^2 + |E_2|^2$) inside the cavity when (a) $d_b=7$ nm and $N=10^{15}$ cm$^{-3}$, (b) $d_b=7$ nm and $N=5\times10^{15}$ cm$^{-3}$, (c) $d_b=3$ nm and $N = 10^{15}$ cm$^{-3}$, and (d) $d_b=3$ nm and $N = 5\times10^{15}$ cm$^{-3}$. In each case, $d_a=7$ nm.}\label{Fig_8}
\end{figure}
%
%-------------------------------------------------------------------------
%
%\subsubsection*{Cavity light intensity}
In Fig.~8, we show time evolution of light intensity $|E_1|^2+|E_2|^2$ inside the laser cavity during the 200th cavity round-trip time when (a) $d_b=7$ nm and $N=10^{15}$ cm$^{-3}$, (b) $d_b=7$ nm and $N=5\times 10^{15}$ cm$^{-3}$, (c)  $d_b=3$ nm and $N=10^{15}$ cm$^{-3}$, and (d) $d_b=3$ nm and $N=5\times 10^{15}$ cm$^{-3}$. In each case, $d_a=7$ nm. For both values of $d_b$, the light intensity significantly increases when $N$ increases from $10^{15}$ cm$^{-3}$ to $5\times 10^{15}$ cm$^{-3}$. Against a red background, a light color represents a bright pulse, and a dark color represents a dark pulse. Self-pulsations of cavity light increase with the increase of $N$. We note that the pulse duration is broader when $N=10^{15}$ cm$^{-3}$ than when $N=5\times 10^{15}$ cm$^{-3}$. Light intensity is much greater when $d_b=7$ nm than when $d_b=3$ nm for $N=10^{15}$ cm$^{-3}$. However, light intensity is relatively similar when $d_b=7$ nm and $d_b=3$ nm for $N=5\times 10^{15}$ cm$^{-3}$.
%

%------------------------------------
%-------------------------------------
% Fig.~9
\begin{figure}[htbp]
\centering
\includegraphics[width=4.2in,keepaspectratio]{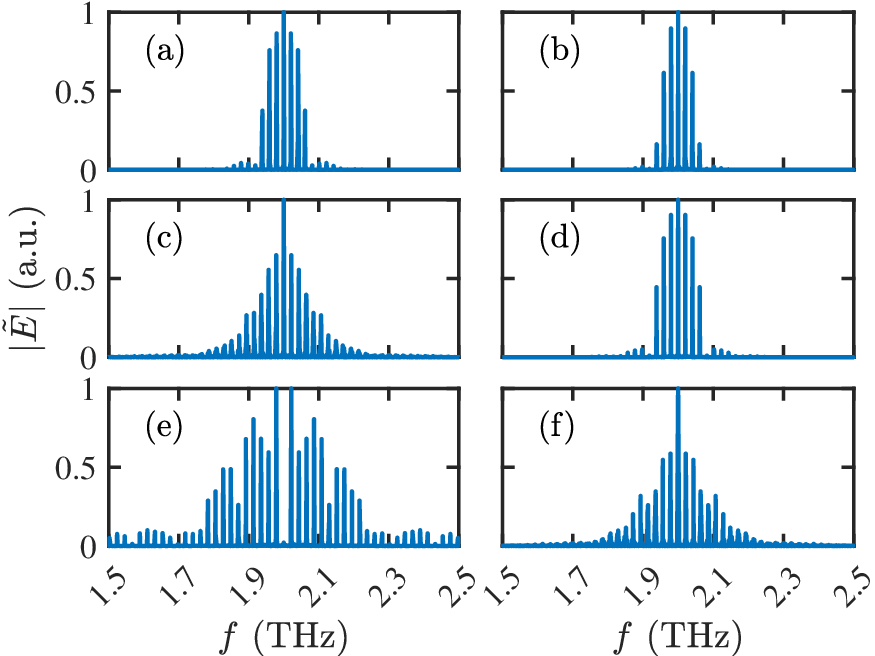}
\caption{Emission spectra ($|\tilde E|$) when $d_b=7$ nm with (a) $N = 10^{15}$ cm$^{-3}$, (c) $N = 2\times 10^{15}$ cm$^{-3}$, and (e) $N = 5\times 10^{15}$ cm$^{-3}$, and when $d_b=3$ nm with (b) $N = 10^{15}$ cm$^{-3}$, (d) $N = 2\times 10^{15}$ cm$^{-3}$, and (f) $N = 5\times 10^{15}$ cm$^{-3}$. In each case, $d_a=7$ nm.}\label{Fig_9}
\end{figure}
%---------------------------------------
%
\subsection{Emission spectra}
This section presents and discusses the emission spectra of the two-photon THz QCL. We note that the intrinsic linewidth of stimulated transitions in the two gain stages is $\Delta\nu=1/(\pi T_2)\approx0.32$ THz. The assumed 0.32 THz intrinsic gain linewidth is typical of THz QCLs. The Fabry-P\'erot cavity for the THz QCL allows only to excite modes with resonances governed by $f=mc/(2nL_C)$, where $m$ is an integer. Therefore, the separation between two axial modes for a 2 mm cavity is $\Delta f=c/(2nL_C)\approx 0.023$ THz.

Figure 9 shows the emission spectra ($|\tilde{E}|$) when $d_a=d_b=7$ nm with (a) $N=10^{15}$ cm$^{-3}$, (c) $N=2\times 10^{15}$ cm$^{-3}$, and (e) $N=5\times 10^{15}$ cm$^{-3}$, and when $d_a=7$ nm and $d_b=3$ nm with (b) $N=10^{15}$ cm$^{-3}$, (d) $N=2\times 10^{15}$ cm$^{-3}$, and (f) $N=5\times 10^{15}$ cm$^{-3}$. We observe that the emission spectra broaden as $N$ increases. Also, the shape of the spectra changes from a close to Gaussian or Lorentzian shape with $N=10^{15}$ cm$^{-3}$ to exponentially decaying with $N=2\times 10^{15}$ cm$^{-3}$ to a modulated spectra with $N=5\times 10^{15}$ cm$^{-3}$ when $d_a=d_b=7$ nm. The emission spectra are significantly broader when $N=5\times 10^{15}$ cm$^{-3}$. We note significant modulation of the emission spectra and splitting into two regions in Fig.~9(e) when $d_a=d_b=7$ nm. This modulation is due to Rabi oscillation, which makes $|\tilde{E}|$ broader with the increase of $N$. 

We note that $|\tilde{E}|$ spectra broaden less with $d_b=3$ nm than $d_b=7$ nm, as shown in Figs.~9(b), (d), and (f). The field per line is also greater when $d_b=7$ nm from its value with $d_b=3$. When $d_b=3$ nm, we do not observe the onset of Rabi splitting when $N=5\times10^{15}$ cm$^{-3}$. Since the two-photon QCL studied emits photons at the same frequency, it is not immediately apparent that the spectra will broaden. However, we observe that the intense emission of two resonant photons increases the number of lasing modes. Furthermore, when Rabi flopping occurs at higher population densities due to the modulation of the emission spectra, the frequency-domain lasing becomes much broader than that of a conventional THz QCL.
%
%------------------------------------------------------------------------------
% Conclusion
\section{Conclusion}
In conclusion, we have presented time- and frequency-resolved FDTD simulation results for a resonant two-photon THz QCL. The FDTD simulation results agree well with the analytical results presented in Ref.~\citeonline{talukder22} and provide an in-depth understanding of two-photon THz QCL dynamics. In the two-photon THz QCL, carrier densities are saturated due to two resonant transitions, and photon-driven scatterings dominate the carrier transport. The time-resolved output intensity is quasi-periodic, and RNGH instability is pronounced when the two-photon nature of the laser is significant. We find the output light intensity increases and emission spectra significantly broaden when the two-photon emission in a THz QCL enhances. The broad emission spectra suggest applications of two-photon THz QCLs in generating frequency combs or where broadband emission is required.
%------------------------------------------------------------------------------
% Reference
\bibliography{qclbib}

\providecommand{\noopsort}[1]{}\providecommand{\singleletter}[1]{#1}%
\begin{thebibliography}{10}

\bibitem{sorokin64}
P.~P. Sorokin and N.~Braslau.
\newblock Some theoretical aspects of a proposed double quantum stimulated emission device.
\newblock {\em IBM Journal of Research and Development}, 8:177--181, 1964.

\bibitem{prokhorov65}
A.~M. Prokhorov.
\newblock Quantum electronics.
\newblock {\em Science}, 149:828--830, 1965.

\bibitem{bunkin66}
F.~V. Bunkin.
\newblock Two quantum transitions in optics.
\newblock {\em Soviet Physics JETP}, 23:1121--1123, 1966.

\bibitem{schlemmer80}
H.~Schlemmer, D.~Fr\"olich, and H.~Welling.
\newblock Two-photon amplification on cascade-transitions.
\newblock {\em Optics Communications}, 32:141--144, 1980.

\bibitem{lewenstein90}
M.~Lewenstein, Y.~Zhu, and T.~W. Mossberg.
\newblock Two-photon gain and lasing in strongly driven two-level atoms.
\newblock {\em Physical Review Letters}, 64:3131--3134, 1990.

\bibitem{zakrzewski91a}
Jakub Zakrzewski, Maciej Lewenstein, and Thomas~W. Mossberg.
\newblock Theory of dressed-state lasers. i. effective hamiltonians and stability properties.
\newblock {\em Physical Review A}, 44:7717--7731, 1991.

\bibitem{zakrzewski91b}
Jakub Zakrzewski, Maciej Lewenstein, and Thomas~W. Mossberg.
\newblock Theory of dressed-state lasers. ii. phase diffusion and squeezing.
\newblock {\em Physical Review A}, 44:7732--7745, 1991.

\bibitem{zakrzewski91c}
Jakub Zakrzewski, Maciej Lewenstein, and Thomas~W. Mossberg.
\newblock Theory of dressed-state lasers. iii. pump-depletion effects.
\newblock {\em Physical Review A}, 44:7746--7758, 1991.

\bibitem{boone89}
A.~W. Boone and S.~Swain.
\newblock Effective hamiltonians and the two-photon laser.
\newblock {\em Quantum Optics: Journal of the European Optical Society Part B}, 1:27--47, 1989.

\bibitem{boone90}
A.~W. Boone and S.~Swain.
\newblock Theory of the degenerate two-photon laser.
\newblock {\em Physical Review A}, 41:343--351, 1990.

\bibitem{yuen78}
H.~Yuen and J.~Shapiro.
\newblock Optical communication with two-photon coherent states--part i: Quantum-state propagation and quantum-noise.
\newblock {\em IEEE Transactions on Information Theory}, 24:657--668, 1978.

\bibitem{pfister01}
O.~Pfister, W.~J. Brown, M.~D. Stenner, and D.~J. Gauthier.
\newblock Polarization instabilities in a two-photon laser.
\newblock {\em Physical Review Letters}, 86:4512--4514, 2001.

\bibitem{brune87b}
M.~Brune, J.~M. Raimond, P.~Goy, L.~Davidovich, and S.~Haroche.
\newblock Realization of a two-photon maser oscillator.
\newblock {\em Physical Review Letters}, 59:1899--1902, 1987.

\bibitem{gauthier92}
D.~J. Gauthier, Q.~Wu, S.~E. Morin, and T.~W. Mossberg.
\newblock Realization of a continuous-wave, two-photon optical laser.
\newblock {\em Physical Review Letters}, 68:464--467, 1992.

\bibitem{ironside92}
C.~N. Ironside.
\newblock Two-photon gain semiconductor amplifier.
\newblock {\em IEEE J. Quantum Electron.}, 28:842--847, 1992.

\bibitem{marti03}
D.~H. Marti, M.-A. Dupertuis, and B.~Deveaud.
\newblock Feasibility study for degenerate two-photon gain in a semiconductor microcavity.
\newblock {\em IEEE J. Quant. Electron.}, 39:1066--1073, 2003.

\bibitem{ning04}
C.~Z. Ning.
\newblock Two-photon lasers based on intersubband transitions in semiconductor quantum wells.
\newblock {\em Physical Review Letters}, 93:187403, 2004.

\bibitem{hayat08}
Alex Hayat, Pavel Ginzburg, and Meir Orenstein.
\newblock Observation of two-photon emission from semiconductors.
\newblock {\em Nature Photonics}, 2:238--241, 2008.

\bibitem{faist94}
Jerome Faist, Federico Capasso, Deborah~L. Sivco, Carlo Sirtori, Albert~L. Hutchinson, and Alfred~Y. Cho.
\newblock Quantum cascade laser.
\newblock {\em Science}, 264:553--556, 1994.

\bibitem{botez23}
Dan Botez and Mikhail~A. Belkin.
\newblock {\em Mid-Infrared and Terahertz Quantum Cascade Lasers}.
\newblock Cambridge University Press, 2023.

\bibitem{vitiello22}
Miriam~Serena Vitiello and Paolo~De Natale.
\newblock Terahertz quantum cascade lasers as enabling quantum technology.
\newblock {\em Advanced Quantum Technologies}, 5:2100082, 2022.

\bibitem{kohler02}
R\"udeger K\"ohler, Alessandro Tredicucci, Fabio Beltram, Harvey~E. Beere, Edmund~H. Linfield, A.~Giles Davies, David~A. Ritchie, Rita~C. Iotti, and Fausto Rossi.
\newblock Terahertz semiconductor-heterostructure laser.
\newblock {\em Nature}, 417:156--159, 2002.

\bibitem{williams07}
Benjamin~S. Williams.
\newblock Terahertz quantum-cascade lasers.
\newblock {\em Nature Photonics}, 1:517--525, 2007.

\bibitem{kumar11}
Sushil Kumar.
\newblock Recent progress in terahertz quantum cascade lasers.
\newblock {\em IEEE Journal of Selected Topics in Quantum Electronics}, 17:38--47, 2011.

\bibitem{burghoff14}
David Burghoff, Tsung-Yu Kao, Ningren Han, Chun Wang~Ivan Chan, Xiaowei Cai, Yang Yang, Darren~J. Hayton, Jian-Rong Gao, John~L. Reno, and Qing Hu.
\newblock Terahertz laser frequency combs.
\newblock {\em Nature Photonics}, 8:462--467, 2014.

\bibitem{barbieri11}
Stefano Barbieri, Marco Ravaro, Pierre Gellie, Giorgio Santarelli, Christophe Manquest, Carlo Sirtori, Suraj~P. Khanna, Edmund~H. Linfield, and A.~Giles Davies.
\newblock Coherent sampling of active mode-locked terahertz quantum cascade lasers and frequency synthesis.
\newblock {\em Nature Photonics}, 5:306--313, 2011.

\bibitem{wang17}
Feihu Wang, Hanond Nong, Tobias Fobbe, Valentino Pistore, Sarah Houver, Sergej Markmann, Nathan Jukam, Maria Amanti, Carlo Sirtori, Souad Moumdji, Raffaele Colombelli, Lianhe Li, Edmund Linfield, Giles Davies, Juliette Mangeney, J\'er\^{o}me Tignon, and Sukhdeep Dhillon.
\newblock Short terahertz pulse generation from a dispersion compensated modelocked semiconductor laser.
\newblock {\em Lasers \& Photonics Review}, 11:1700013, 2017.

\bibitem{talukder22}
Muhammad~Anisuzzaman Talukder, Paul Dean, Edmund~Harold Linfield, and Giles Davies.
\newblock Resonant two-photon terahertz quantum cascade laser.
\newblock {\em Optics Express}, 30:31785--31794, 2022.

\bibitem{valcarcel95}
G.~J. de~Valc\'arcel and Eugenio Rold\'an.
\newblock Two-photon laser dynamics.
\newblock {\em Physical Review A}, 52:4059--4069, 1995.

\bibitem{talukder14}
Muhammad~Anisuzzaman Talukder and Curtis~R. Menyuk.
\newblock Quantum coherent saturable absorption for mid-infrared ultra-short pulses.
\newblock {\em Optics Express}, 1:15608--15617, 2014.

\bibitem{talukder09apl}
Muhammad~Anisuzzaman Talukder and Curtis~R. Menyuk.
\newblock Effects of backward-propagating waves and lumped mirror losses on self-induced transparency modelocking in quantum cascade lasers.
\newblock {\em Applied Physics Letters}, 95:071109, 2009.

\bibitem{talukder09pra}
Muhammad~Anisuzzaman Talukder and Curtis~R. Menyuk.
\newblock Analytical and computational study of self-induced transparency mode locking in quantum cascade lasers.
\newblock {\em Physical Review A}, 79:063841, 2009.

\bibitem{williams03}
Benjamin~S. Williams, Hans Callebaut, Sushil Kumar, Qing Hua, and John~L. Reno.
\newblock 3.4-thz quantum cascade laser based on longitudinal-optical-phonon scattering for depopulation.
\newblock {\em Applied Physics Letters}, 82:1015--1017, 2003.

\bibitem{kumar08}
Sushil Kumar and Alan W.~M. Lee.
\newblock Resonant-phonon terahertz quantum-cascade lasers and video-rate terahertz imaging.
\newblock {\em IEEE Journal of Selected Topics in Quantum Electronics}, 14:333--344, 2008.

\bibitem{liu12}
Tao Liu, Tillmann Kubis, Qi~Jie Wang, and Gerhard Klimeck.
\newblock Design of three-well indirect pumping terahertz quantum cascade lasers for high optical gain based on nonequilibrium green's function analysis.
\newblock {\em Applied Physics Letters}, 100:122110, 2012.

\bibitem{risken68}
H.~Risken and K.~Nummedal.
\newblock Self-pulsing in lasers.
\newblock {\em Journal of Applied Physics}, 39:4662--4672, 1968.

\bibitem{gordon08}
Ariel Gordon, Christine~Y. Wang, L.~Diehl, F.~X. K\"artner, A.~Belyanin, D.~Bour, S.~Corzine, G.~H\"ofler, H.~C. Liu, H.~Schneider, T.~Maier, M.~Troccoli, J.~Faist, and Federico Capasso.
\newblock Multimode regimes in quantum cascade lasers: From coherent instabilities to spatial hole burning.
\newblock {\em Physical Review A}, 77:053804, 2008.

\bibitem{talukder09prl}
Curtis~R. Menyuk and Muhammad~Anisuzzaman Talukder.
\newblock Self-induced transparency modelocking of quantum cascade lasers.
\newblock {\em Physical Review Letters}, 102:023903, 2009.

\end{thebibliography}
\bibliographystyle{unsrt}
\end{document}